\documentclass{revtex4}
\begin{document}
\title{The Gutzwiller wave function as a disentanglement prescription}
\author{D.K.~Sunko\thanks{email: dks\@phy.hr}}
\affiliation{Department of Physics,\\Faculty of Science,\\
University of Zagreb,\\
Bijeni\v cka cesta 32,\\
HR-10000 Zagreb, Croatia.}
\date{Oct. 3, 2004}

\begin{abstract}
The Gutzwiller variational wave function is shown to correspond to a
particular disentanglement of the thermal evolution operator, and to be
physically consistent only in the temperature range $U\ll kT\ll E_F$, the
Fermi energy of the non-interacting system. The correspondence is established
without using the Gutzwiller approximation. It provides a systematic
procedure for extending the ansatz to the strong-coupling regime. This is
carried out to infinite order in a dominant class of commutators. The
calculation shows that the classical idea of suppressing double occupation
is replaced at low temperatures by a quantum RVB-like condition, which
involves phases at neighboring sites. Low-energy phenomenologies are
discussed in the light of this result.
\end{abstract}

\maketitle

\section{Introduction}

Variational wave functions are a highly select class of results in the
physics literature. There are only five which are widely used:
Hartree-Fock and BCS for weak perturbations of the Fermi
sea~\cite{Mahan90}, Feynman's ansatz for the ground state of liquid
$^4$He~\cite{Feynman54}, Laughlin's wave function for the fractional
quantum Hall effect~\cite{Laughlin83}, and Gutzwiller's ansatz for the
ground state of the Hubbard model~\cite{Gutzwiller65}. The last two fall
into the class of Jastrow wave functions~\cite{Jastrow55}, one of which
was also used to describe $^4$He and $^4$He--$^3$He
mixtures~\cite{McMillan68}, and which are currently considered generic for
strongly correlated problems.

Of these, Gutzwiller's is by far the least understood. Its underlying
physical idea is that electrons of one spin see those of the other as a
`smeared background'~\cite{Gutzwiller65}. This very drastic assumption is
still not sufficient to provide an operational prescription, but is
supplemented by another, the `Gutzwiller approximation:' electrons of one
spin see the others `as if occupying a band of width
zero'~\cite{Gutzwiller65}, \emph{i.e.} their mass is taken to be infinite.
This prescription was never given a justification from first principles in
any finite dimension. It is true by construction when the number of
dimensions approaches infinity~\cite{Metzner89}, because the scaling of
hopping overlaps, required to obtain finite results in that limit, makes
all motion effectively diffusive.

The present work approaches Gutzwiller's wave function from a perspective not
suggested by its variational origin. It turns out that it is based on a
one-step Trotter decomposition of the thermal evolution operator, strictly
valid only if the on-site repulsion $U$ is much lower than the temperature.
This insight provides a natural scheme for improvement. A direct
implementation of it shows that Gutzwiller's prescription to remove double
occupation is the first step in a transcedent series. When summed, it yields
a new projector, which imposes a quantum condition with much stronger
selectivity than the one removing double occupation. In the physical subspace
satisfying this condition, Gutzwiller's program may be carried over to the
strong-coupling regime $kT\ll t\ll U$ as well, where $t$ is the hopping
overlap. It can also be shown that at least at the level of expectation
values, the on-site interaction does not scatter out of the new physical
subspace. Unlike the requirement of no double occupancy, the quantum
condition cannot be factorized into commuting local terms, indicating that
relative phases on neighboring sites play an important role in the
realization of the insulating ground state. The arguments are limited to the
immediate vicinity of half-filling, where the configuration space for
processes not considered here is small.

\section{The Gutzwiller ansatz}\label{mapping}

Take the Hubbard Hamiltonian on a square lattice, $H=K+V$, where $K$ is
the kinetic term and $V=U\sum_in_{i\uparrow}n_{i\downarrow}$. Define the
operator ${\mathcal{K}}$ by a factorization of the imaginary-time
evolution operator,
\begin{equation}
e^{-\beta(K+V)}\equiv e^{-\beta V/2}e^{-\beta\mathcal{K}}e^{-\beta V/2}.
\label{exfac}
\end{equation}
The main result of this section is that the Gutzwiller ansatz neglects the
entanglement of $K$ and $V$. To prove this, take $\mathcal{K}=K$ and
calculate the expectation with respect to any operator $\mathcal{O}$:
\begin{eqnarray}
\frac{
\mathrm{tr}\; \mathcal{O}e^{-\beta V/2}e^{-\beta K}e^{-\beta V/2}
}{\mathrm{tr}\; e^{-\beta K}}
&=&
\frac{
\sum_{P}\left<P\right|e^{-\beta V/2}\mathcal{O}e^{-\beta V/2}e^{-\beta K}
\left|P\right>
}{\mathrm{tr}\; e^{-\beta K}}
\nonumber\\
&=&
\sum_{PRR'}\left<P|R\right>\left<R\right|
e^{-\beta V/2}
\mathcal{O}
e^{-\beta V/2}\left|R'\right>
\left<R'|P\right>
\frac{
\left<P\right|e^{-\beta K}\left|P\right>
}{\mathrm{tr}\; e^{-\beta K}}
\nonumber\\&=&
\sum_{RR'}\left[
e^{-\beta U(D_R+D_{R'})/2}
\left<R\right|\mathcal{O}\left|R'\right>\right]
\sum_{P}
\frac{
\left<P\right|e^{-\beta K}\left|P\right>
}{\mathrm{tr}\; e^{-\beta K}}
\left<R'|P\right>
\left<P|R\right>.
\label{expec}
\end{eqnarray}
Here $\left|P\right>$ are momentum eigenstates, and
$\left|R\right>,\left|R'\right>$ position eigenstates. Use has been made
of the fact that $K$ is diagonal in momentum, and $V$ in position:
$V\left|R\right>=UD_R\left|R\right>$, where $D_R$ is the number of doubly
occupied sites in configuration $\left|R\right>$. Now perform the same
calculation for the expectation value
$\left<g\right|\mathcal{O}\left|g\right>$ in Gutzwiller's wave function
$\left|g\right>=P(g)\left|\Psi\right>$, with $\left|\Psi\right>$ a
non-interacting ground state and $P(g)$ the Gutzwiller projector:
\begin{equation}
P(g)=\prod_i\left[
1-(1-g)n_{i\uparrow}n_{i\downarrow}
\right]\equiv\prod_i\left[
\widehat E_i+\widehat A_{i\uparrow}+\widehat A_{i\downarrow}+g\widehat D_i
\right],
\label{gutz}
\end{equation}
where the hatted operators are projectors onto empty sites, sites occupied
by a single spin (up or down), and doubly occupied sites,
respectively~\cite{Ogawa75}. Then
\begin{eqnarray}
\left<g\right|\mathcal{O}\left|g\right>
&=&
\sum_{RR'}\left<\Psi|R\right>\left<R\right|
P(g)
\mathcal{O}
P(g)\left|R'\right>\left<R'\right.
\left|\Psi\right>\nonumber\\
&=&
\sum_{RR'}\left[
g^{D_R+D_{R'}}
\left<R\right|\mathcal{O}\left|R'\right>\right]
\left<R'|\Psi\right>
\left<\Psi|R\right>,
\label{expecgutz}
\end{eqnarray}
remembering that $P(g)\left|R\right>=g^{D_R}\left|R\right>$. Now observe that
the non-interacting ground state $\left|\Psi\right>$ is itself a momentum
eigenstate. The same term $P=\Psi$ will dominate the sum over $P$ in
Eq.~(\ref{expec}), if one takes the temperature low enough.  The expectation
value~(\ref{expec}) then reads
\begin{equation}
\mathrm{tr}\; \mathcal{O}e^{-\beta^* V/2}e^{-\beta^* K}e^{-\beta^* V/2}
\approx
\sum_{RR'}\left[
e^{-\beta^* U(D_R+D_{R'})/2}
\left<R\right|\mathcal{O}\left|R'\right>\right]
\left<R'|\Psi\right>\left<\Psi|R\right>,
\label{expec0}
\end{equation}
where $\beta^*$ is a particular value of the temperature, for which the
non-interacting system is in its ground state, to any desired accuracy.

We are led to the astonishing conclusion, that the result~(\ref{expec0}) of
this procedure is the same as taking expectation values with respect to
Gutzwiller's variational wave function, Eq.~(\ref{expecgutz}). The
correspondence
\begin{equation}
\left<g\right|\mathcal{O}\left|g\right>
\leftrightarrow
\mathrm{tr}\; \mathcal{O}e^{-\beta^* V/2}e^{-\beta^* K}e^{-\beta^* V/2}
\end{equation}
between the two expressions~(\ref{expecgutz}) and~(\ref{expec0}) is
established simply by replacing
\begin{equation}
g\leftrightarrow e^{-\beta^* U/2}.
\label{gcorr}
\end{equation}
The fact that the denominator in Eq.~(\ref{expec}) was $\mathrm{tr}\;
e^{-\beta K}$ and not (more logically) $\mathrm{tr}\; e^{-\beta K}e^{-\beta
V}$ is counterpart to the fact that $\left<g|g\right>$ is not normalized.
Taking $\mathcal{O}=1$, it immediately follows that
$\left<g|g\right>\leftrightarrow  \mathrm{tr}\; e^{-\beta K}e^{-\beta
V}/\mathrm{tr}\;e^{-\beta K}$ under the above correspondence.

The original calculation of Gutzwiller also contains a prescription to fix
$\beta^*$, or $g$. This is to take $\mathcal{O}=H$, the Hamiltonian, and
obtain $g$ variationally. However, it is difficult to imagine such a
procedure to compensate for the steps which were taken to arrive at
Gutzwiller's form, Eq.~(\ref{expec0}). Neglecting entanglement to get from
Eq.~(1) to Eq.~(2) requires $U\ll kT$, or more precisely, $Ut\ll (kT)^2$.
This is the step usually made in the Trotter formula, for a single short
`slice' of the evolution integral, which is eventually taken to zero
(\emph{i.e.} the temperature to infinity). To single out the ground-state
term in Eq.~(2) and so obtain Eq.~(\ref{expec0}), requires, on the other
hand, the temperature to be low, $kT\ll E_F$, the Fermi energy of the
non-interacting ground state. The two are possibly consistent only in the
range
\begin{equation}
U\ll kT\ll E_F,
\end{equation}
which is not the strong-coupling regime $kT\ll t\ll U$, for which the
Gutzwiller approach was intended.

On the other hand, in practice the expectation value~(\ref{expecgutz}) is
usually calculated in the so-called Gutzwiller approximation, so it is
desirable to understand its effect on the above derivation. It was shown in
Ref.~\cite{Ogawa75} that the approximation amounted to replacing the
configurational overlaps by a constant,
\begin{equation}
\left<R'|P\right>\left<P|R\right>
\rightarrow\frac{1}{\mathcal{N}},
\end{equation}
where $\mathcal{N}$ is the number of terms in the sum over configurations
$R,R'$. Inserting this in Eq.~(\ref{expec}) yields
\begin{equation}
\frac{1}{\mathcal{N}}
\sum_{RR'}\left[
e^{-\beta^* U(D_R+D_{R'})/2}
\left<R\right|\mathcal{O}\left|R'\right>\right],
\end{equation}
which is the same as obtained from Eq.~(\ref{expecgutz}) with Gutzwiller's
approximation, without invoking the limit $kT\ll E_F$. The physical role of
the Gutzwiller approximation is clear now. Instead of neglecting the
\emph{excited} states in Eq.~(\ref{expec}), as done above by going to low
temperature, it neglects the \emph{difference} between the ground and excited
states. When all the terms $\left<R'|P\right>\left<P|R\right>$ are replaced
by the constant $1/\mathcal{N}$, of course their thermal average in
Eq.~(\ref{expec}) reduces to this single constant term. One gets again the
same result as if the ground-state term alone had been taken into account.

To summarize, calculating the expectation value of any operator with respect
to Gutzwiller's wave function is exactly equivalent to the following three
steps, when calculating its thermal expectation with respect to
$e^{-\beta(K+V)}$: first (a) neglect the entanglement of the potential
and kinetic terms, so that one can replace
\begin{equation}
e^{-\beta(K+V)}\rightarrow e^{-\beta V/2}e^{-\beta K}e^{-\beta V/2},
\label{disent}
\end{equation}
forgetting all commutators of $V$ and $K$; then (b) take only the
ground-state term from Eq.~(\ref{expec}); finally (c) use a variational
procedure to fix the left-over temperature-dependent parameter $e^{-\beta^*
U/2}=g$, irrespective of consistency with the previous two steps. Needless to
say, the first two steps are themselves hardly justifiable in the
strong-coupling limit $U\gg t$.

\section{Lowest-order improvement}

In the previous section, it was shown that the expectation values
calculated with the Gutzwiller ansatz can be obtained in a thermal
formalism which neglects the entanglement of kinetic and potential energy
terms. This is true irrespective of the use of the Gutzwiller
approximation, which itself amounts to replacing the thermal average in
Eq.~(\ref{expec}) with a normalization constant. The whole procedure
retains only a classical attenuation of double occupation, such that
Eq.~(\ref{exfac}) is approximately valid with $\mathcal{K}=K$, \emph{i.e.}
all quantum dynamical correlations, coming from the commutators, are
neglected. The main subject of the present work is to investigate the
effect of the neglected commutators systematically.

Including the commutators amounts to adding quantum correlations to
Gutzwiller's wave function, which should be present in the strong-coupling
low-temperature state, $kT\ll t< U$. Technically, this boils down to finding
a better expression for the operator $\mathcal{K}$ in Eq.~(\ref{exfac}). The
reason ${\mathcal K}\neq K$ is that the commutator $[V,K]$ is not zero.
Explicitly,
\begin{equation}
V^k\circ K=tU^k\sum_{<i,j>\atop\sigma}
\left(n_{i,-\sigma}-n_{j,-\sigma}\right)^k
\left( a^{\dagger}_{i\sigma} a^{\vphantom{\dagger}}_{j\sigma}+(-1)^k
a^{\dagger}_{j\sigma} a^{\vphantom{\dagger}}_{i\sigma}\right),
\label{polcom}
\end{equation}
where $t$ is the hopping overlap, and the operation $\circ$ is a commutator,
\begin{equation}
V^n\circ K\equiv \left[V,V^{n-1}\circ K\right]
\label{comm}
\end{equation}
with $V^0\circ K\equiv K$. The vanishing of the commutator is obviously
consistent with the original `smeared background' interpretation,
$n_{i,-\sigma}\to\left<n_{-\sigma}\right>$. This points the way to an \emph{a
posteriori} justification of the Gutzwiller ansatz (though not of the
Gutzwiller approximation). One can claim to work in a physical regime where
it is sensible to replace the number operators by the average occupation of a
site, as it should be in a doped metallic state, away from the
metal-insulator transition. There one can hope that the high-temperature
decomposition, Eq.~(\ref{disent}), may in fact extend to low temperature.

In the remainder of this section, the effect of including a single additional
commutator will be studied. It will be shown below that the commutator $[V,K]$
itself does not contribute to $\mathcal{K}$, because of the symmetry of the
decomposition~(\ref{exfac}), so the lowest non-zero contributions to same
order in $\beta$ are $[V,[V,K]]$ and $[K,[V,K]]$. When $U\gg t$, the limit of
interest here, the first is more important. Retaining only this one term,
\begin{eqnarray}
\mathcal{K}
&\rightarrow&
K+\frac{1}{6}\left(\frac{\beta}{2}\right)^2[V,[V,K]]
\nonumber\\&=&
t\sum_{<i,j>\atop\sigma}
\left[1+\frac{(\beta U)^2}{24}n_{ij,-\sigma}
\right]
\left( a^{\dagger}_{i\sigma} a^{\vphantom{\dagger}}_{j\sigma}+
a^{\dagger}_{j\sigma} a^{\vphantom{\dagger}}_{i\sigma}\right)\equiv K+K_1,
\label{kprim}
\end{eqnarray}
where the numerical factors will also be justified later, and
\begin{equation}
n_{ij,-\sigma}=
\left(n_{i,-\sigma}-n_{j,-\sigma}\right)^2=
n_{i,-\sigma}+n_{j,-\sigma}-2n_{i,-\sigma}n_{j,-\sigma}
\end{equation}
is equal to one if the hop changes the number of doubly occupied sites, and
zero otherwise.

How can one use this result to improve Gutzwiller's ansatz? Note that the norm
of Gutzwiller's wave function can be written
\begin{equation}
\left<\Psi\right|P(g)P(g)\left|\Psi\right>\leftrightarrow
\frac{
\mathrm{tr}\; e^{-\beta V/2}e^{-\beta K}e^{-\beta V/2}
}{\mathrm{tr}\; e^{-\beta K}}
\label{proj0}
\end{equation}
under the formal correspondence of the previous section. Obviously, one can
interpret this as
\begin{equation}
P(g)\leftrightarrow
e^{-\beta V/2}.
\label{projcorr}
\end{equation}
To confirm the interpretation, recall the well-known alternative
form~\cite{Fulde93} of writing the Gutzwiller projector~(\ref{gutz}),
\begin{equation}
P(g)=\exp\left[-\eta\sum_in_{i\uparrow}n_{i\downarrow}
\right]=g^{\sum n_{i\uparrow}n_{i\downarrow}},
\end{equation}
where $\eta=-\ln g$. But this is just the right-hand side
of~(\ref{projcorr}), under the correspondence~(\ref{gcorr}).

Now, the additional commutator in Eq.~(\ref{kprim}) amounts to replacing $K$
by $K+K_1$ in the numerator of~(\ref{proj0}). Since we have decided not to
include any additional commutators, we may rearrange terms at will, and write
\begin{equation}
\frac{
\mathrm{tr}\; e^{-\beta V/2}e^{-\beta K_1/2}
e^{-\beta K}
e^{-\beta K_1/2}e^{-\beta V/2}
}{\mathrm{tr}\; e^{-\beta K}}
\label{proj1}
\end{equation}
for the `improved' right-hand side of~(\ref{proj0}). It is obvious how to
write an improved left-hand side now. There should be an additional projector,
sensitive to configurations in which a hop would change the number of
doubly occupied sites. Calling it $P_1(g_1)$, one may write
\begin{equation}
P_1(g_1)\leftrightarrow
e^{-\beta K_1/2}.
\end{equation}
Explicitly,
\begin{equation}
P_1(g_1)=\exp\left[-\eta_1\sum_{<i,j>\atop\sigma}
n_{ij,-\sigma}\left( a^{\dagger}_{i\sigma} a^{\vphantom{\dagger}}_{j\sigma}+
a^{\dagger}_{j\sigma} a^{\vphantom{\dagger}}_{i\sigma}\right)
\right]=
g_1^{\sum
n_{ij,-\sigma}\left( a^{\dagger}_{i\sigma} a^{\vphantom{\dagger}}_{j\sigma}+
a^{\dagger}_{j\sigma} a^{\vphantom{\dagger}}_{i\sigma}\right)},
\label{newproj}
\end{equation}
where $\eta_1=-\ln g_1$, and the new variational parameter corresponds to
\begin{equation}
g_1\leftrightarrow e^{-\beta^3 tU^2/48}.
\label{g1corr}
\end{equation}
Hence the (systematically) improved variational wave function is
\begin{equation}
\left|g,g_1\right>=P_1(g_1)P(g)\left|\Psi\right>.
\label{improved}
\end{equation}
The practical aspects of evaluating the projector~(\ref{newproj}) are beyond
this article. One immediately obvious technical complication is, however,
theoretically significant: the projector~(\ref{newproj}) cannot be written in
the product form~(\ref{gutz}). The reason is formally that the individual
terms in the sum in Eq.~(\ref{newproj}) do not commute. Physically, this
means that even the first quantum correction to Gutzwiller's program for the
one-band Hubbard model already involves the relative phases of fermions on
neighboring sites. The importance of phases was already noticed in
Ref.~\cite{Millis91}, where it was shown that if a projector \emph{can} be
factorized into commuting local terms, then it \emph{cannot} produce physical
insulating behavior. In the following, it will be argued that while the use
of projectors to remove unwanted parts of configuration space may be
perfectly valid, such projectors cannot be used automatically to define
effective Hamiltonians.

\section{Extension to all orders}

In the previous section, it was shown that the first quantum correction to
Gutzwiller's projection involves bond phases. This is a setback for the local
approach, so one is naturally led to ask, how important that correction is.
What is the range of validity, in temperature, of the original ansatz? As
luck would have it, this question admits of a sharp answer, because the
correlation embodied in $K_1$ (or P$_1$) above can easily be studied to all
orders in $\beta$. Unfortunately, the answer is rather disappointing: the
low-temperature regime is approached exponentially fast. This statement
should be moderated insofar as we really have in mind the state near
half-filling. As already noted, any argument establishing a `smeared
background' automatically justifies Gutzwiller's ansatz at all temperatures.
On the other hand, in the vicinity of the metal-insulator transition, one
expects an electron to be scattered many times for each step it takes: there
is not enough propagation in space to average out the local density. Then
(\ref{polcom}) contains the dominant processes affecting kinetic motion near
half-filling, when $U> t\gg kT$.

The resummation of all commutators in~(\ref{polcom}) does not introduce
new correlations, because
\begin{equation}
\left(n_{i,-\sigma}-n_{j,-\sigma}\right)^{2k}=
\left(n_{i,-\sigma}-n_{j,-\sigma}\right)^2 = n_{ij,-\sigma}
\label{projec}
\end{equation}
is independent of $k$. However, it will show how fast (in temperature)
they become important, if one can calculate how these commutators enter
$\mathcal{K}$. The linear terms in a Baker-Campbell-Hausdorff (BCH)
formula can be obtained by standard tricks~\cite{Bourbaki75}, to give
\begin{equation}
{\mathcal{K}}=\frac{\sinh\left({\beta V\over 2}\right)}{{\beta V\over 2}}
\circ K
+ \left(1-{\beta K\over 2}\coth{\beta K\over 2}\right)\circ V
+ \mathcal{O}(\beta^4K^2V^2)
\label{bch}
\end{equation}
for the case of Eq.~(\ref{exfac}). (Some details are given in
Appendix~\ref{Bourbaki}.) The first term is now evaluated with the help of
Eq.~(\ref{polcom}), remembering that the projector~(\ref{projec}) may be
taken outside the Taylor series. One obtains
\begin{equation}
{\mathcal{K}}=t\sum_{<i,j>\atop\sigma}
\left[1+s(\beta U)n_{ij,-\sigma}\right]
\left( a^{\dagger}_{i\sigma} a^{\vphantom{\dagger}}_{j\sigma}+
a^{\dagger}_{j\sigma} a^{\vphantom{\dagger}}_{i\sigma}\right)
+\mathcal{O}(\beta^2K^2)\equiv K+K_1+\mathcal{O}(\beta^2K^2).
\label{correct}
\end{equation}
where we have redefined $K_1$. The point is now that the function
\begin{equation}
s(\beta U)=\frac{\sinh\left({\beta U\over 2}\right)}{{\beta U\over 2}}-1
\end{equation}
grows exponentially with $\beta U$. The interpretation~(\ref{g1corr}) of the
variational parameter $g_1$ should be replaced by
\begin{equation}
g_1\leftrightarrow e^{-\beta t\,s(\beta U)/2}.
\label{g1fin}
\end{equation}
The projector $P_1$ in Eq.~(\ref{improved}) becomes important at least as
soon as
\begin{equation}
s(\beta U)>1,
\label{cond}
\end{equation}
when $K_1$ and $K$ become competitive. This is a much sharper condition than
intuitively expected. On the other hand, it is also non-linear, so the
correction in~(\ref{correct}) is only 4\% for $kT=U$, and the
condition~(\ref{cond}) is first satisfied for $kT\approx U/4.4$~. One may thus
replace the lower limit of validity of Gutzwiller's ansatz by $U/4\ll kT$
(say), but that is obviously not essential.

Comparing the interpretations~(\ref{gcorr}) of Gutzwiller's parameter,
and~(\ref{g1fin}) of $g_1$, more can be said. When $s(\beta U)> U/t$, the new
projector $P_1$ becomes more important than Gutzwiller's $P$. Thus the 
very-low-temperature regime $kT\ll U$ is completely dominated by $P_1$,
unless $t/U$ is exponentially small, which is not normally the case. (The
limit $U/t\to\infty$ is discussed later on.) For $kT\ll t<U$, the
wave-function $P_1(g_1)\left|\Psi\right>$ represents reality much better
(exponentially better, to coin a phrase) than $P(g)\left|\Psi\right>$. The
strong-coupling limit is denoted $t<U$ rather than $t\ll U$ to avoid
confusion with the limit $U/t\to\infty$, because we need $s(\beta U)\gg U/t$
when $kT\ll U$. However, $U/t$ is always taken to be sufficiently large to
relegate the neglected commutators to weak perturbations. Note that, since
$s(\beta U)$ rises exponentially, such a regime is easily achieved. For
example, for $kT=t/10\ll t< U=10t$, one gets
$$
s(\beta U)=s(100)\sim 10^{19}\gg U/t=10.
$$

The formal reason for the overwhelming dominance of $P_1$ at low temperature
is that the iterated commutator of Hubbard's contact interaction is non-zero
to all orders. By comparison, for the harmonic oscillator already the third
iterated commutator vanishes: $[\hat{x}^2,[\hat{x}^2,\hat{p}^2]]\sim
\hat{x}^2$, and similarly when $\hat{x}$ and $\hat{p}$ are interchanged. If
all possible commutators are arranged in a table, such that the $(n,m)$ cell
collects those of order $V^nK^m$, this table is tridiagonal for the harmonic
oscillator, while the expression~(\ref{polcom}) gives the first column for
the contact interaction.

The final question in this section is about the relevance of the other
commutators in the above-mentioned table, of which so far only the first
column was treated. At $k$-th order in $\beta$, commutators contribute which
are of order $K^mV^{k-m}\sim t^mU^{k-m}$. The first column, summed completely
by the hyperbolic sine in Eq.~(\ref{bch}), contains all terms with $m=1$,
\emph{i.e.} like $tU^{k-1}$, while the other columns refer respectively to
$1<m<k$. Thus at any given order in $\beta$, the term contained in the
hyperbolic sine can be made to dominate those left out simply by increasing
$U/t$. However, one should not hastily conclude that projectors generated by
these terms are dominated by $P_1$ at low temperature in the same sense as
Gutzwiller's $P(g)$. Like $P_1$, they are expected to have hyperbolic terms
in $\beta U$ in the exponent, where $P$ is only linear in $\beta U$. Thus
they should be relatively mildly suppressed with respect to $P_1$, by a power
of the ratio $t/U$. However, since suppression by $P_1$ is the strongest, the
configuration space determined by this projector alone is the largest
one that needs to be taken into account, at least in the vicinity of
half-filling.

\section{The physical subspace}

Looking back at the wave function~(\ref{improved}), in the light of the
underlying dependence~(\ref{g1fin}) of $g_1$ with temperature, one may well
wonder: what will survive such a projector at low temperature? Even if a wave
function has a very small component to which $K_1$ in~(\ref{correct}) is
sensitive, the hyperexponential suppression by $P_1$ will
annihilate it. The way out is obvious: the physically admissible space is
the null-space of the operator $K_1$, $K_1\left|\Phi\right>=0$, or
equivalently
\begin{equation}
[V,[V,K]]\left|\Phi\right>=tU^2\sum_{<i,j>\atop\sigma}
n_{ij,-\sigma}\left( a^{\dagger}_{i\sigma} a^{\vphantom{\dagger}}_{j\sigma}+
a^{\dagger}_{j\sigma} a^{\vphantom{\dagger}}_{i\sigma}\right)
\left|\Phi\right>=0.
\label{null}
\end{equation}
This gives a precise meaning to the `smeared background' condition at low
temperature and near half-filling, where it cannot be trivially satisfied by
the replacement $n_{ij,-\sigma}\to 0$. A more careful formulation of the same
idea is
\begin{equation}
\exp\left[-\beta(K+K_1)\right]\left|\Phi\right>=
\exp\left(-\beta K\right)\left|\Phi\right>.
\label{K1}
\end{equation}
(The first of these conditions implies the second, but not vice versa.)
However, $K_1$ dominates $K$ at low temperature because of the relative
factor $s(\beta U)$, so one expects the physics to be contained in
Eq.~(\ref{null}) by itself. One may be tempted to add the condition
\begin{equation}
V\left|\Phi\right>=0,
\label{nodub}
\end{equation}
which is the no-double-occupancy constraint, but that is not warranted: for
example, should the conditions~(\ref{null}) and~(\ref{nodub}) turn out to be
incompatible, the discussion in the previous section shows that the system
will choose~(\ref{null}) at low temperature. By the same token, the
subdominant correlations, coming from the neglected commutators, may have a
say in which combination of the states $\left|\Phi\right>$ turns out to be
the ground state, but they cannot enlarge the physical subspace any more than
the no-double occupancy constraint: $P_1$ acts too stringently for that (if
it does not, decrease the temperature, and/or increase the ratio $U/t$). If
we decide to neglect all commutators containing at least two $K$'s, such as
$[K,[V,K]]$, because they are suppressed by at least $t/U$, the thermal
expectation value in the physical subspace may be written at low temperature
\begin{equation}
\mathrm{tr}\; \mathcal{O}e^{-\beta V/2}e^{-\beta K+K_1}e^{-\beta V/2}
=
\sum_{\Phi}\left<\Phi\right|e^{-\beta V/2}\mathcal{O}e^{-\beta V/2}e^{-\beta K}
\left|\Phi\right>,
\label{phys}
\end{equation}
because the wave-functions which do not satisfy~(\ref{null}) have been
eliminated by the projector $P_1$, while the admissible ones allow the
simplification~(\ref{K1}). This has formally the same structure as if we
had made the high-temperature disentanglement~(\ref{disent}), \emph{i.e.}
used Gutzwiller's scheme. However, the underlying physical regime is
at low temperature and strong coupling, ensured by the requirement~(\ref{null})
on the states $\left|\Phi\right>$.

If $U/t\to\infty$, the dominance of $P_1$ over Gutzwiller's $P$ cannot be
established quantitatively. This does not invalidate the former reasoning,
but merely opens the possibility that the physical subspace is further
reduced in the calculation~(\ref{phys}). However, there is a qualitative
argument that Hubbard's interaction leaves the physical subspace invariant.
Namely, the identity
\begin{equation}
VAV=\frac{1}{2}\left(V^2A+AV^2-[V,[V,A]]\right)
\end{equation}
holds for any operators $V$ and $A$. Now take $V$ to be Hubbard's
repulsion, and $A=[V,[V,K]]$, \emph{cf.} Eq.~(\ref{null}). Let
$\left|\Phi\right>$ be a state satisfying Eq.~(\ref{null}), \emph{i.e.}
$A\left|\Phi\right>=0$. Then the state $V\left|\Phi\right>$ also satisfies
Eq.~(\ref{null}), at the level of expectation values:
\begin{equation}
\left<\Phi\right|VAV\left|\Phi\right>=
-\frac{1}{2}\left<\Phi\right|[V,[V,A]]\left|\Phi\right>\propto
\left<\Phi\right|A\left|\Phi\right>=0,
\end{equation}
because $[V,[V,A]]\propto A$, by virtue of~(\ref{polcom})
and~(\ref{projec}). In other words, Hubbard's repulsion does not on the
average scatter out of the null-space defined by Eq.~(\ref{null}). This
is strong indication that the latter has been correctly identified as the
physical subspace.

\section{Effective Hamiltonians}

In the previous section, the physical subspace~(\ref{null}) was found to have
two important properties. One, it is on the average invariant to $V$. Two,
within it Gutzwiller's disentanglement (`smeared background') holds, so that
hopping proceeds by the bare operator $K$, \emph{cf.} Eq.~(\ref{K1}). This
should be contrasted with the no-double-occupancy subspace, which is strongly
perturbed by $K$, requiring the introduction of projected hopping
\begin{equation}
\widetilde{K}=t\sum_{<i,j>\atop\sigma}
\left(1-n_{i,-\sigma}\right)a^{\dagger}_{i\sigma}
a^{\vphantom{\dagger}}_{j\sigma}\left(1-n_{j,-\sigma}\right)+
\mathit{h.c.},
\end{equation}
to keep within it. Both schemes take into account that an electron cannot
hop locally onto anything but an empty site, in which case it tunnels by
the full overlap $t$. In the first one, the burden of accounting for these
dynamical correlations is taken by the construction~(\ref{null}) of the
physical subspace, while in the no-double-occupancy scheme it is carried by
the projected-hopping operator $\widetilde{K}$.

To compare the two clearly, note that using the projected-hopping operator in
the no-double-occupancy subspace amounts to the calculation of thermal traces
of the type
\begin{equation}
\mathrm{tr}\; \mathcal{O}e^{-\beta V/2}e^{-\beta\widetilde{K}}e^{-\beta
V/2}
\end{equation}
at arbitrarily low temperature. Now, the \emph{exact} expression for the
trace is
\begin{equation}
\mathrm{tr}\; \mathcal{O}e^{-\beta (K+V)}
=
\mathrm{tr}\; \mathcal{O}e^{-\beta V/2}e^{-\beta\mathcal{K}}e^{-\beta V/2},
\end{equation}
by the definition~(\ref{exfac}) of the operator $\mathcal{K}$. As shown in
the previous sections, the relevant part of $\mathcal{K}$ at low
temperature is $K+K_1$, which in turn reduces either to $K_1$ in the
unphysical subspace (since $K_1\gg K$ there), or to $K$ in the physical
subspace, defined to be the null-space of $K_1$. Even if this separation
into physical and unphysical subspaces turned out to be wholly misguided,
still the fact would remain, that $\mathcal{K}$ has no component of the
type $\widetilde{K}$. The neglected commutators cannot change this,
because (a) they can be independently suppressed, simply by increasing the
ratio $U/t$, and (b) they contain $K$ at least twice, as in $[K,[V,K]]$,
so they generate three-site and spin-exchange correlations, which are
absent in $\widetilde{K}$.

In fact even more can be said: neither the classical
constraint~(\ref{nodub}), nor the quantum constraint~(\ref{null}) can be used
to define an effective Hamiltonian. The reason is that they are both
\emph{negative} statements, excluding some unwanted correlations. An
effective Hamiltonian, on the other hand, acts to build up desirable
correlations, not to reject undesirable ones. For example, the Hartree-Fock
Hamiltonian has the respective Slater determinant as its ground state. By
contrast, the null-space conditions~(\ref{null}) or~(\ref{nodub}) give no
indication, which combination of states satisfying them is quasi-stationary
with respect to the Hubbard Hamiltonian. Having an effective Hamiltonian is
equivalent to knowing the approximate (saddle-point) equation of motion
within the physical subspace, which is much more than knowing which states
are not in that subspace.

There is a simple and rather dramatic way to emphasize that null-space
conditions are not equations of motion --- or, equivalently, that projectors
are not effective Hamiltonians. Within its regime of validity, one should be
able to use an effective Hamiltonian $H_{\mathit{eff}}$ just as if it were
fundamental, \emph{i.e.} forgetting that its elementary degrees of freedom
have an internal structure. In particular, the same $H_{\mathit{eff}}$ should
be used in real and imaginary time. This reflects the requirement that the
effective equations of motion admit the ensemble construction, \emph{i.e.}
are thermalized in the usual sense. Now, what is the real-time analogue of
$K+K_1$, Eq.~(\ref{correct})? It is found by taking $\beta=i\tau$ in the
function $s(\beta U)$,
\begin{equation}
s(i\tau U)=\frac{\sin\left({\tau U\over 2}\right)}{{\tau U\over 2}}-1.
\end{equation}
Note the different dependence on $U$. To drive the point home,
consider the limit $\tau U\to\infty$. Then $s\to -1$, and
\begin{eqnarray}
K+K_1&\to&t\sum_{<i,j>\atop\sigma}
\left[(1-n_{i,-\sigma}) a^{\dagger}_{i\sigma}
a^{\vphantom{\dagger}}_{j\sigma}(1-n_{j,-\sigma})+{\mathit (h.c.)}\right]
\nonumber\\
&&+\;t\sum_{<i,j>\atop\sigma}\left[n_{i,-\sigma} a^{\dagger}_{i\sigma} 
a^{\vphantom{\dagger}}_{j\sigma}n_{j,-\sigma}+{\mathit (h.c.)}\right]
\label{keff}
\end{eqnarray}
because the projector which appears may be rewritten:
$$
1+s(i\tau U)\,n_{ij,-\sigma}\to
1-n_{ij,-\sigma}=(1-n_{i,-\sigma})(1-n_{j,-\sigma})+
n_{i,-\sigma}n_{j,-\sigma}.
$$
The first term in Eq.~(\ref{keff}) is just the projected-hopping operator
$\widetilde{K}$, while the second makes the whole expression particle-hole
symmetric. The appearance of $\widetilde{K}$ under these conditions is
direct evidence that it is not an effective Hamiltonian for the one-band
Hubbard model. If it were, it could be used in imaginary time as well,
while it was shown above that the correct imaginary-time expression is
Eq.~(\ref{correct}), not $\widetilde{K}$.

The formal way out of this real/imaginary time conundrum is again to work
in the subspace~(\ref{null}), since the kinetic operator then reverts to
the microscopic (bare) $K$ on both axes. However, that is not the same as
having an effective Hamiltonian. The prescription~(\ref{phys}) correctly
transcribes Gutzwiller's calculational framework to the regime $kT\ll
t<U$, but this is at best the first step in finding the normal modes,
still far from revealing them.

\section{Discussion}

The standard operator approach to an effective theory without double
occupation is to find a similarity transformation~\cite{Fulde93,Harris67}
\begin{equation}
SHS^{-1}=H_{\mathit{eff}}+\ldots,
\label{fw}
\end{equation}
such that $H_{\mathit{eff}}$ does not couple to the doubly occupied states
to some given order, and truncate the remainder, represented above by
dots. The present paper observes that such a procedure does not of itself
guarantee a physical effective Hamiltonian. In particular, it is shown
that avoiding double occupation is a semiclassical perception of the
electron's behavior at strong coupling, which does not carry over to low
temperature near half-filling. Imposing null-space conditions does not
imply there exist thermodynamically stable normal modes which satisfy
them. If the approximate normal modes are known, it may indeed be possible
to connect effective Hamiltonians with low-lying states written in
projector form, as in the cases of Hartree-Fock and BCS. If they are not,
using the connection by formal analogy runs into trouble, such as not
having the same effective Hamiltonian on the real and imaginary time axes.
In other words, while an equation of motion may be written as a null-space
condition, $(H-E)\left|\Psi\right>=0$, not all null-space conditions are
admissible constraints to some given equations of motion. Put more simply,
fixing both the force and the effective constraint due to that force
generally amounts to overspecifying the problem. To be certain the two
are compatible is almost the same as knowing the solution.

The projected-hopping operator $\widetilde{K}$ is a case in point, since it
appears in the literature in two different contexts. On the one hand, it may
be derived by formal arguments based on~(\ref{fw}), guided by the wish to
avoid some parts of configuration space. On the other, the same
$\widetilde{K}$ can be obtained by physical arguments from a more general
three-band model, where it is claimed to describe the propagation of
excitations against the background of a particular mode, the Zhang-Rice
singlet~\cite{Zhang88}. In this case, the use of $\widetilde{K}$ as a true
effective Hamiltonian depends only on whether the correct hierarchy of
background, excitation, and thermalization time scales is established. The
point is that one cannot derive an effective Hamiltonian without some image
of a physical mode which becomes quasi-stationary under the action of the
original one. The same is true in the variational context. For example, in
his description of $^3$He--$^4$He mixtures~\cite{McMillan68}, McMillan
checked that the Jastrow wave function for the $^4$He background reduced in
the long-wavelength limit to the density oscillations characterizing
Feynman's formulation~\cite{Feynman54}.

There is nothing wrong, in principle, with using projector language to
guess properties of the solution, \emph{i.e.} to try and delimit the
physical subspace. This is the essence of Gutzwiller's program, which the
present work expands systematically. It is shown here that the original
program is (at best) consistent in the physical regime $U\ll kT\ll E_F$,
and that it may be carried over to the regime $kT\ll t<U$ formally very
simply, by working in the null-space of the `most troublesome' commutator,
Eq.~(\ref{null}). This was identified with the physical subspace in the
low-temperature, strong coupling limit, by a projection argument: states
which are not in this null-space were found to be hyperexponentially
suppressed at low temperature, much more strongly than states containing
double occupation. The question may of course be raised, whether keeping
the simplification of Gutzwiller's disentanglement at low temperature is
worth the price of working with the condition~(\ref{null}). In effect, the
preservation of disentanglement has replaced avoiding double occupation as
a guiding principle for the construction of the physical subspace. This
requirement is at least consistent between real and imaginary time, but it
remains to be seen whether it is compatible as a constraint with the
microscopic on-site repulsion. Physically, it amounts to the conjecture
that low-lying excitations in the Hubbard model near half-filling can be
mapped onto an effective semiclassical gas. It is encouraging for this
point of view that the on-site repulsion does not on the average scatter
out of the new physical subspace.

The condition~(\ref{null}) is the first precise quantum formulation of
Gutzwiller's `smeared background' assumption. Its most interesting aspect
is the role of phases. Indeed the expression indicates that the admissible
physical states should be coherent, as suggested by the
resonating-valence-bond (RVB) arguments of Anderson~\cite{Anderson87}.
This should be contrasted with the viewpoint, based on the
no-double-occupancy condition~(\ref{nodub}), that the system is insulating
because the electron \emph{locally} has difficulty overcoming the
repulsion $U$. The latter leads to an essentially diffusive view of the
Mott state, which was shown already in Ref.~\cite{Millis91} not to give
physical insulating behavior, precisely because it is incoherent,
\emph{i.e.} insensitive to relative phases on neighboring sites. The fact
that the interaction loses coherence, $[K,V]=0$, as soon as
$n_{ij,\sigma}=0$, then implies that the ground-state of the one-band
Hubbard model is likely to be antiferromagnetic at half-filling, even at
arbitrarily large $U/t$. This conjecture is specific to the one-band
model, since $n_{ij,\sigma}$ appeares in the commutator only because both
sites, connected by the hopping, are subject to the local repulsion.

While this article was under review, new evidence appeared that the
no-double-occupancy condition is not satisfied at low temperature in the
one-band Hubbard model~\cite{Phillips04}. It turns out that the upper
Hubbard band participates coherently in the low-energy density of states
at half-filling, even for $U/t$ as large as 12.

To conclude, it has been shown that the Gutzwiller variational ansatz is
physically consistent only in the regime $U\ll kT\ll E_F$, inapplicable to
the Mott transition. In the regime $U>t\gg kT$, the no-double-occupancy
condition is replaced by a quantum condition~(\ref{null}), sensitive to local
phases, which defines the physical subspace. Within this subspace,
Gutzwiller's disentanglement scheme is recovered in the low-temperature,
strong coupling limit as well. However, as always, one cannot find an
effective Hamiltonian without knowing the dominant slow modes. The quantum
projection~(\ref{null}) is hopefully a step forward in understanding their
microscopic structure, but is not an equation of motion.

\acknowledgments

Conversations with S.~Bari\v si\'c and E.~Tuti\v s are gratefully
acknowledged. Thanks are due to D.~Svrtan for helping with
Ref.~\cite{Bourbaki75}, and to P.~W.~Anderson for pointing out
Ref.~\cite{McMillan68}. This work was supported by the Croatian Government
under Project~0119256.

\appendix

\section{Derivation of the first term in Equation~(\ref{bch})}\label{Bourbaki}

Let $A$ and $B$ be two algebraic indeterminates. Define
\begin{equation}
X=\ln\left(e^{A+B}e^{-A}\right),\,Y=\ln\left(e^{-A}e^{A+B}\right).
\end{equation}
It may be shown~\cite{Bourbaki75} that the parts linear in $B$ of these
expressions are
\begin{equation}
X\equiv_{\mbox{\tiny B}}{e^A-1\over A}\circ B=X_B,\,Y\equiv_{\mbox{\tiny B}}
{1-e^{-A}\over A}\circ B=Y_B,
\label{bour}
\end{equation}
where the notation $\equiv_{\mbox{\tiny B}}$ means `equal up to terms
linear in B,' and the circle operation is a commutator, like in
(\ref{comm}). The linear parts are called $X_B$ and $Y_B$, as noted. Now
\begin{equation}
\ln\left(e^{-A}e^{2A+2B}e^{-A}\right)=\ln\left(e^Ye^X\right)
\equiv_{\mbox{\tiny B}}\ln\left(e^{Y_B}e^{X_B}\right)
\equiv_{\mbox{\tiny B}}X_B+Y_B,
\end{equation}
where the second step is legal because neither $X$ nor $Y$ contain terms of
order zero in $B$, and the third is trivial, since both $X_B$ and $Y_B$ are
linear in $B$. Adding $X_B$ and $Y_B$, one obtains
\begin{equation}
\ln\left(e^{-A}e^{2A+2B}e^{-A}\right)
\equiv_{\mbox{\tiny B}}{e^A-e^{-A}\over A}\circ B.
\end{equation}
This gives the first term in Eq.~(\ref{bch}), putting $A=-\beta V/2$ and
$B=-\beta K/2$.

For completness, here is the derivation of (\ref{bour}) from
Ref.~\cite{Bourbaki75}. First, introduce a convenient notation for left-
and right-multiplication by $A$,
\begin{equation}
LB=AB,\,RB=BA,
\end{equation}
so that, for instance, the commutator with $A$ is written $(L-R)B$. Then one
can write $A^pBA^q=L^pR^qB$, whence it is trivial to show
\begin{equation}
e^ABe^{-A}=e^Le^{-R}B=e^{L-R}B=e^A\circ B,
\label{eAB}
\end{equation}
reverting to the circle notation. In the same vein,
\begin{equation}
(A+B)^n\equiv_{\mbox{\tiny B}} A^n+\sum_{k=0}^{n-1} A^kBA^{n-1-k}=
A^n+\sum_{k=0}^{n-1} L^kR^{n-1-k}B,
\end{equation}
so that the factorization formula
\begin{equation}
(L-R)\sum_{k=0}^{n-1} L^kR^{n-1-k}=L^n-R^n
\end{equation}
yields
\begin{equation}
A\circ (A+B)^n\equiv_{\mbox{\tiny B}} (L^n-R^n)B=A^nB-BA^n,
\end{equation}
from which follows the useful expression
\begin{equation}
A\circ e^{A+B}\equiv_{\mbox{\tiny B}}e^A B-Be^A,
\label{AeAB}
\end{equation}
upon summation over $n$.

To get (\ref{bour}), first note that $X\equiv_{\mbox{\tiny B}} X_B$ (no
zeroth-order term), hence $e^X\equiv_{\mbox{\tiny B}} 1+X_B$. Then
\begin{eqnarray}
\lefteqn{
A\circ X_B\equiv_{\mbox{\tiny B}}Ae^X-e^XA=Ae^{A+B}e^{-A}-e^{A+B}e^{-A}A}
\nonumber\\
&=&\left(Ae^{A+B}-e^{A+B}A\right)e^{-A}\equiv_{\mbox{\tiny B}}
\left(e^A B-Be^A\right)e^{-A}=
e^ABe^{-A}-B=\left(e^A-1\right)\circ B,
\end{eqnarray}
where (\ref{AeAB}) and (\ref{eAB}) were used in succesion. This means that
in order to get $X_B$ itself, we need one less commutator in each term on
the right-hand side. In the circle notation, this is just
\begin{equation}
X_B={e^A-1\over A}\circ B,
\end{equation}
which is the first expression in (\ref{bour}). The second is obtained in
exactly parallel fashion.

\end{document}